\documentclass[amsmath,pbl, elsarticle, twocolumn,prl] {revtex4}
\usepackage{bm}
\usepackage{enumerate}
\usepackage{graphicx}
\usepackage[dvips]{epsfig}
\usepackage{epsf}
\usepackage{xcolor}
\makeatletter

\newcommand{\Rmnum}[1]{\expandafter\@slowromancap\romannumeral #1@}

\makeatletter

\begin{document}
\title{d-Wave Hall effect and linear magnetoconductivity in metallic collinear antiferromagnets}

\author{Dmitrii L. Vorobev}
\affiliation{Higher School of Economics University, 101000 Moscow, Russia}
\affiliation{L.D. Landau Institute for Theoretical Physics, 142432, Chernogolovka, Russia}

\author{Vladimir A. Zyuzin}
\affiliation{L.D. Landau Institute for Theoretical Physics, 142432, Chernogolovka, Russia}

\begin{abstract}
In this paper we theoretically predict a distinct class of anomalous Hall effects occurring in metallic collinear antiferromagnets. 
The effect is quadratic and $d-$wave symmetric in external magnetic field. 
In addition, the electric current, transverse to the current voltage drop and the magnetic field in the predicted effect are all in the same plane.
Studied theoretical model consists of two-dimensional fermions interacting with the N\'{e}el order through momentum dependent exchange interaction having a $d-$wave symmetry.
We also find unusual linear magnetoconductivity in this model.
\end{abstract}

\maketitle

Anomalous Hall effect (AHE) is one of the experimental tools which sheds light on the symmetries of the crystal sctructure of the material.
The effect primarily requires time-reversal symmetry breaking. 
The details of the crystal structure and magnetic order can vary the AHE.
For example, in ferromagnets with magnetization ${\bf M}$ (or Zeeman part of external magnetic field ${\bf B}$) it is expected \cite{KarplusLuttinger} there will be
\begin{align}\label{AHE}
{\bf j}_{\mathrm{AHE}} \propto \left[ {\bf M} \times {\bf E} \right],
\end{align}
electric current response, where ${\bf E}$ is the electric field corresponding to the voltage drop.
This is the ferromagnetic analog of the regular Hall effect \cite{Hall,Ziman}. Main ingredients of AHE in ferromagnets is the combination of the exchange (momentum independent) spin splitting and the spin-orbit coupling \cite{Vas'ko,BychkovRashba,Dresselhaus,SovietTI1,SovietTI2,Dyakonov} of the conducting fermions.  For example, two-dimensional fermion system with Rashba spin-orbit coupling and Zeeman-like ferromagnetic exchange is one of the most studied models in relation to the anomalous Hall effect \cite{CulcerMacDonaldNiuPRB2003,AHE_RMP,SinitsynPRB2007,NunnerPRB2007}. 
One of the main mechnisms of the AHE is the anomalous part of the fermion velocity \cite{KarplusLuttinger} which on the other hand is due to the Berry curvature \cite{Berry,BerryReview} of conducting fermions.

It is then possible that in systems with C$_{1v}$ and C$_{3v}$ symmetry the magnetic field driven AHE can have in-plane configuration \cite{Malshukov1998,LiuPRL2013,ZhangPRB2019,Zyuzin2020,Culcer2021,Kurumaji2023}, where all three vectors, namely the electric current, transverse voltage drop and external magnetic field, are in the same, in the example below $x-y$, plane,
\begin{align}\label{IPHE}
{\bf j}_{\mathrm{IPHE}} = \sigma_{\mathrm{IPHE}} \left[ \left[{\bf B}\times {\bf e}_{z}  \right] \times {\bf e}_{y}\right]\times {\bf E},
\end{align}
here unit vectors ${\bf e}_{z}$ and ${\bf e}_{y}$ (in $z-$ and $y-$ directions correspondingly) are defined by the spin-orbit coupling (see \cite{Zyuzin2020} for details), and $\sigma_{\mathrm{IPHE}}$ is the in-plane Hall conductivity.
First two vector products filter out only the $B_{y}$ component multiplied by the ${\bf e}_{z}$ unit vector.
This effect has been experimentally observed in Refs. \cite{AHE_ZrTe5, Zhang_IPHE2020,Liu2024_C3v}. In addition, \cite{LiuPRL2013,ZhangPRB2019,Zyuzin2020} suggested that in systems with C$_{3v}$ symmetry $\sigma_{\mathrm{IPHE}} \propto B_{y}^2 - 3B_{x}^2$ vanishing at $B_{y}=\pm\sqrt{3}B_{x}$ overall making ${\bf j}_{\mathrm{IPHE}}$ current $\frac{2\pi}{3}$ periodic in the in-plane magnetic field, which was experimentally observed in \cite{Liu2024_C3v}.

\begin{figure}[t] 
\centerline{
\begin{tabular}{cc}
\includegraphics[width=0.5 \columnwidth]{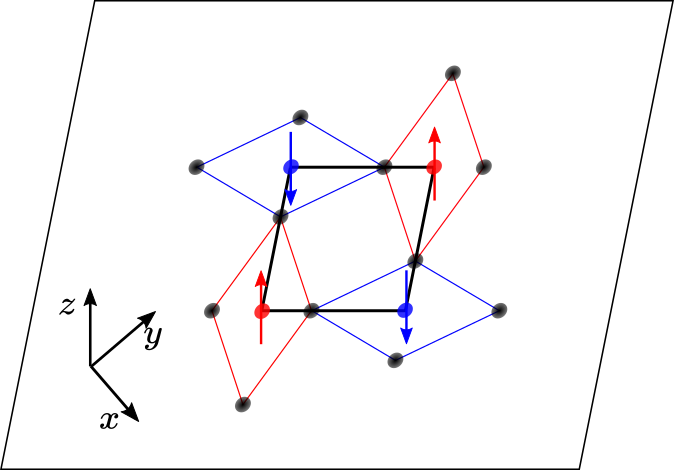}~~
\includegraphics[width=0.35 \columnwidth]{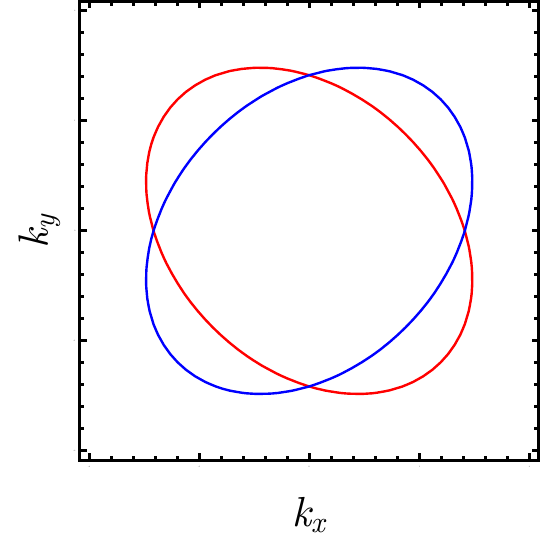}
\end{tabular}
}
\protect\caption{Left: schematics of two-dimensional square lattice in $x-y$ plane with N\'{e}el order in $z-$ direction. 
Red/blue circles correspond to two atoms in the unit cell with spin up/down order. 
Local environment composed of black circles is different around red/blue atoms.
There is a symmetry of $\frac{\pi}{2}$ rotation together with time-reversal:
the $\frac{\pi}{2}$ rotation connects local environment of the two elements while the time-reversal flips the spin. 
Right: Fermi surface of fermions corresponding to Hamiltonian Eq. (\ref{Hamiltonian}) with $\lambda = 0$. 
Red corresponds to spin up, while blue to spin down.}

\label{fig:fig1}  

\end{figure}

The situation with AHE in collinear antiferromagnets is currently under research \cite{AHE_AFM_Review,AHE_AFM}.
In simple collinear antiferromagnets, with two sublattices, there is a symmetry under a combination of time-reversal and translation operations which does not allow for spin-splitting of conducting fermions, therefore, making the anomalous Hall effect to vanish.
However, microscopic surroundings of each sublattice \cite{AHE_AFM} may make a difference. 
For example, the aforementioned symmetry will be broken if surroundings of spin up sublattice is different from the surroundings of the spin down. See left plot in Fig. (\ref{fig:fig1}) for schematics. 
The remaining symmetry is a combination of time-reversal and $\frac{\pi}{2}$ rotation operations, which will allow for the spin-splitting of conducting fermions shown in the right plot in Fig. (\ref{fig:fig1}). 
In the level of Hamiltonian, such spin-splitting can be understood as the momentum-dependent $g-$ factor which obeys symmetry of the lattice. 
Research of materials with momentum-dependent $g-$factor has been studied for decades \cite{PekarRashba,Ogg1966,IvchenkoKiselevPTS1992,VarmaZhu2006,WuSunFradkinZhang2007, Ramazashvili,HayamiYanagiKusunose2019}. However, the existence of such a $g-$factor in antiferromagnetic materials has been pointed out only recently \cite{AHE_AFM,Rashba2020,HayamiYanagiKusunose2020,EgorovEvarestov,SmejkalSinovaJungwirth2022a, SmejkalSinovaJungwirth2022b, Bose2022,Gonzalez-Hernandez2021,Exp2023}.  
It is understood that conducting fermions in, for example, RuO$_{2}$, MnF$_{2}$, FeSb$_{2}$, MnTe N\'{e}el ordered antiferromagnets and many more can be described by such spin splitting \cite{AHE_AFM,Rashba2020,EgorovEvarestov,SmejkalSinovaJungwirth2022a, SmejkalSinovaJungwirth2022b}. 
Experimentally measured \cite{Bose2022} spin filtered electric transport is one of the manifestations of such spin splitting \cite{Gonzalez-Hernandez2021}. We note that there are other proposals \cite{Guo_npj2023} on how asymmetry of the sublattices in N\'{e}el ordered antiferromagnets may result in a non-zero AHE.

In this paper we show that in addition to known cases of the anomalous Hall effect, given by Eq. (\ref{AHE}) and (\ref{IPHE}), 
metallic antiferromagnets with spin split conducting fermions described above may show very unusual magnetic field driven anomalous Hall effect propoprtional to the second power of the magnetic field, given by
\begin{align}\label{dwave}
{\bf j}_{\mathrm{DWHE}} \propto   B_{x}B_{y}\left[{\bf e}_{z}\times {\bf E}\right],
\end{align}
where ${\bf e}_{z}$ is defined by the direction of the N\'{e}el vector. 
We will be referring to it as the $d-$wave Hall effect, hence the DWHE abbreviation in Eq. (\ref{dwave}). Indeed the $B_{x}B_{y}$ product has the aforementioned symmetry.
It is notable that just like in the in-plane Hall effect Eq. (\ref{IPHE}) all three vectors, namely electric current, transverse voltage drop, and the magnetic field, are in the same plane. 
Such a response is not prohibited by the Onsager relation, as it is overall cubic in time-reversal symmetry breaking fields, since ${\bf e}_{z} \rightarrow -{\bf e}_{z}$ in Eq. (\ref{dwave}) under the time-reversal operation.

In addition to (\ref{dwave}) we find another experimentally relevant response, namely the linear magnetoconductivity (LMC),
\begin{align}\label{lmc}
{\bf j}_{\mathrm{LMC}} \propto  B_{z}\left(E_{x}{\bf e}_{y} + E_{y}{\bf e}_{x} \right),
\end{align}
which, together with regular Hall effect, will result in anisotropic Hall conductivity, i.e. $\sigma_{xy} \neq \sigma_{yx}$. 
Again, this effect is allowed by Onsager relation because response in (\ref{lmc}) is actually quadratic in time-reversal symmetry breaking fields, since $B_{z}$ is selected by N\'{e}el order ${\bf e}_{z}$ as $B_{z} \rightarrow ({\bf e}_{z}\cdot{\bf B})$ and both change sign under time-reversal.

%--------------------------------------------------------------------------------------------------------------------------------------------------------------------
%--------------------------------------------------------------------------------------------------------------------------------------------------------------------

In order to show the effect, we study a two-dimensional metallic antiferromagnet system shown in left plot in Fig. (\ref{fig:fig1}). 
The Hamiltonian of conducting fermions interacting with the N\'{e}el vector and consistent with the lattice symmetry is written as
\begin{align}\label{Hamiltonian}
\hat{H}_{0} = \frac{{\bf k}^2}{2m} 
+ \lambda (k_{x}\sigma_{y} - k_{y}\sigma_{x})
+ \beta \sigma_{z}k_{x}k_{y},
\end{align}
where $\sigma$ are Pauli matrices describing spin of conducting fermions. 
The term with $\beta$ is the interaction of the conducting fermions with the antiferromagnetic N\'{e}el vector (coined as the altermagnetism by \cite{AHE_AFM,SmejkalSinovaJungwirth2022a, SmejkalSinovaJungwirth2022b}). 
This term breaks the time-reversal symmetry.
As noted, it is a combination of difference in atomic configurations around ordered spins and the antiferromagnetic order which generates this term, not just the antiferromagnetic order alone. 
For example, see left plot in Fig. (\ref{fig:fig1}), where a combination of translation and time-reversal symmetries is broken due to the local configuration, while a $\frac{\pi}{2}$ rotation together with the time-reversal is the symmetry of the lattice which supports the term $\beta$.
We have included Rashba spin-orbit coupling denoted by $\lambda$. In addition we apply external magnetic field in $x-y$ plane which acts only on spins of fermions (Zeeman magnetic field),
\begin{align}\label{Z}
\hat{H}_{\mathrm{Z}} = h_{x}\sigma_{x}+h_{y}\sigma_{y},
\end{align}
where ${\bf h} = \frac{1}{2}g\mu_{\mathrm{B}}{\bf B}$, $g$ is the $g-$factor, and $\mu_{\mathrm{B}}$ is the Bohr magneton. Both terms will be needed in our analysis of the electric current responses.
In addition, orbital part of the magnetic field in $z-$direction will be considered.
We assume that finite electron density with chemical potential $\mu >0$ does not suppress the antiferromagnetic order, as well as the external Zeeman magnetic field does not cant the antiferromagnetically ordered spins in any direction. The latter is plausable if the field is smaller than the magnetic anisotropy which favors $z-$ direction for the N\'{e}el vector. 

The Hamiltonian Eq. (\ref{Hamiltonian}) hasn't been derived by us from microscopics, it is an effective model which corresponds to the lattice symmetry \cite{AHE_AFM, SmejkalSinovaJungwirth2022a, SmejkalSinovaJungwirth2022b}. 
However, one may think of it as a model of correlated fermions on a square lattice. 
Then an antiferromagnetic order develops in the vicinity of the half-filling gapping out the fermions. Slight doping with fermions populates the conduction band with quadratic spectrum (first term in Eq. (\ref{Hamiltonian})). 
If in addition there is a different local environment of spins shown in the left figure in Fig (\ref{fig:fig1}), the spectrum of fermions will acquire a term with $\beta$ in Eq. (\ref{Hamiltonian}).   
One may also think of a term with $\beta$ as a $d-$wave Pomeranchuk or Stoner-like magnetic instability \cite{VarmaZhu2006,WuSunFradkinZhang2007}.

We are interested in electric current responses of this system to external electric field. 
Let us first understand what kind of responses can be deduced from the symmetry argument. 
Symmetry group corresponding to the unperturbed system, which is the fermions on a simple square lattice without anything else, is $D_{4\mathrm{h}}$ group. 
The other terms are treated as perturbations, and in this particular symmetry group \cite{Koster} field $\beta$ transforms as $\Gamma_{3}^{(+)}$ element of the group, magnetic field $B_{z}$ as $\Gamma_{2}^{(+)}$ and $B_{x/y}$ as $\Gamma_{5}^{(+)}$, the electric field as $\Gamma_{5}^{(-)}$, $\lambda$ as $\Gamma_{2}^{(-)}$, and the electric current transforms as $\Gamma_{5}^{(-)}$.
The elements of the group obey multiplication rules, for example, listed in \cite{Koster,BirPikus,Snoke}.
We must find all products that are linear in electric field, linear in $\beta$, to second order in the magnetic field, and to whatever order in $\lambda$ that transform as $\Gamma_{5}^{(-)}$ and hence can be a part of electric current. 
In addition we require Onsager relation for the conductivity $\sigma_{ij}({\bf B},\beta) = \sigma_{ji}(-{\bf B},-\beta) $ to satisfy.
By performing excercise of multiplying the group elements we get for the electric current
\begin{align}\label{currentSymmetry}
{\bf j} 
&= \sigma_{\mathrm{D}} {\bf E} + \sigma_{\mathrm{H}} \left[{\bf E}\times {\bf B} \right] + \sigma_{2}\left[ ({\bf E}\cdot {\bf B}){\bf B}  - {\bf B}^2 {\bf E} \right]
\\
&
 + \sigma_{\mathrm{LMC}}\beta B_{z}\left(E_{x}{\bf e}_{y} + E_{y}{\bf e}_{x} \right) + \sigma_{\mathrm{DWHE}}\beta B_{x}B_{y}\left[{\bf e}_{z}\times {\bf E}\right].
\nonumber
\end{align}
The first line here is consistent with Refs. \cite{Ziman,SeitzPR1950,GoldbergDavisPR1954}.
A term with $\sigma_{\mathrm{D}}$ is the regular Drude conductivity, with $\sigma_{\mathrm{H}} =\frac{1}{B} \omega_{\mathrm{c}}\tau \sigma_{\mathrm{D}}$, where $\omega_{\mathrm{c}} = \frac{eB}{mc}$ is the cyclotron frequency, is the regular Hall effect due to the Lorentz force \cite{Hall}, $\sigma_{2}$ is due to the Lorentz force as well (in Weyl semimetals this term can be due to the chiral anomaly, for example see \cite{ZyuzinWSM}, and in ferromagnets it is called as the planar Hall effect if ${\bf B}$ is replaced by the magnetization ${\bf M}$ \cite{KyJETP1966}) and exists in any three-dimensional electron system \cite{Ziman,SeitzPR1950,GoldbergDavisPR1954}, 
$\sigma_{\mathrm{LMC}}$ is the LMC expected in the time-reversal symmetry broken systems \cite{CortijoPRB2016,ZyuzinWSM,Zyuzin2021,comment2,WSMCorrelated,LeeRosenbaum2020,ExpPRL2021}, and finally a term with $\sigma_{\mathrm{DWHE}}$ is the DWHE. Last two terms in Eq. (\ref{currentSymmetry}) are unique to the system described by $\hat{H}_{0} + \hat{H}_{\mathrm{Z}}$. 
Let us demonstrate how they appear (please see Supplemental Material \cite{SM} for more details). According to the multiplication table given in \cite{Koster,BirPikus,Snoke}, $E_{x/y}B_{z}\beta$ transform as $\Gamma_{5}^{(-)} \times \Gamma_{2}^{(+)} \times \Gamma_{3}^{(+)} = \Gamma_{5}^{(-)}$ while $E_{x/y}B_{x}B_{y}\beta$ as $\Gamma_{5}^{(-)} \times \Gamma_{5}^{(+)} \times \Gamma_{5}^{(+)} \times \Gamma_{3}^{(+)} = \Gamma_{5}^{(-)}$. Indeed, the two combinations transform as electric current.
In addition to the Lorentz force contribution to $\sigma_{\mathrm{H}}$ there might be a contribution from regular anomalous Hall effect given by Eq. (\ref{AHE}) if ${\bf M}$ there is replaced by $B_{z}{\bf e}_{z}$.

If mechinism behind each term in the first line of Eq. (\ref{currentSymmetry}) is understood \cite{Ziman}, terms in the second line haven't been discussed anywhere before and are subjects of the analysis below.
We first introduce the notations. The spectrum for $s=\pm$ branches corresponding to $\hat{H}_{0} + \hat{H}_{\mathrm{Z}}$ Hamiltonian reads, 
\begin{align}\label{spectrum}
\varepsilon_{\bf k}^{(\pm)} = \frac{k^2}{2m} \pm \sqrt{\Delta_{\bf k}^2  + \lambda^2 \tilde{k}^2  } ,
\end{align}
where $\lambda \tilde{k}_{x} = \lambda k_{x} + h_{y}$ and $\lambda\tilde{k}_{y} =\lambda k_{y} - h_{x}$ and $\Delta_{\bf k} = \beta k_{x}k_{y}$ were introduced for brevity. 
Spinors are
$
\Psi_{{\bf k},+} = \left[
 \cos\left( \frac{\xi_{\bf k}}{2}\right) e^{i\chi_{\bf k}} ,
 - \sin\left( \frac{\xi_{\bf k}}{2}\right) \right]^{\mathrm{T}} 
 $
 and
 $
\Psi_{{\bf k},-} = \left[
 \sin\left( \frac{\xi_{\bf k}}{2}\right) e^{i\chi_{\bf k}},
 \cos\left( \frac{\xi_{\bf k}}{2}\right)  \right]^{\mathrm{T}},
$
where $[..]^{\mathrm{T}}$ is the transposition, $\cos(\xi_{\bf k}) = \frac{\Delta_{\bf k}}{   \sqrt{ \Delta_{\bf k}^2 + \lambda^2 \tilde{k}^2 }}$, and $\chi_{\bf k} = \arctan\left(\frac{\tilde{k}_{y} }{\tilde{k}_{x}}\right)$ is the phase.

\begin{figure}[h] 
\centerline{
\includegraphics[width=0.9 \columnwidth]{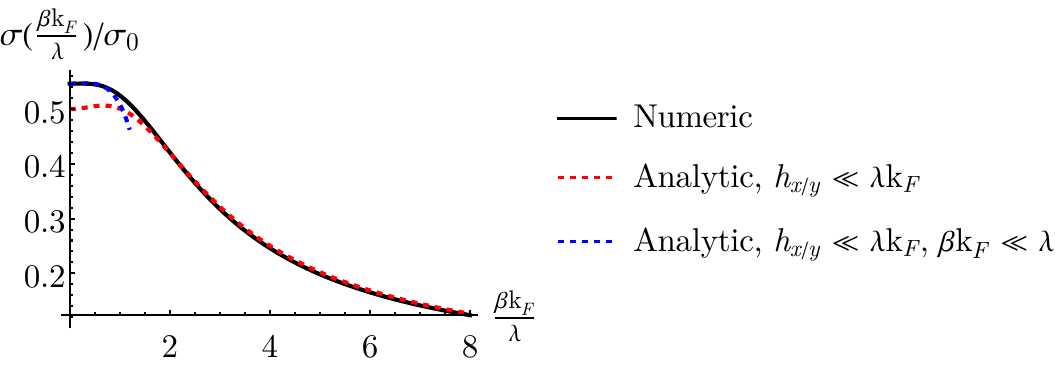} 
}
\protect\caption{
Plot of the d-wave Hall conductivity $\sigma_{\mathrm{DWHE}}$ given by Eq. (\ref{eq1}) and Eq. (\ref{eq2}). Values of the parameters used in the numerical calculation are $\frac{h_{x}}{\lambda k_{\mathrm{F}}}=0.25$, $\frac{h_{y}}{\lambda k_{\mathrm{F}}}=0.15$, and $T=0$. We have defined $\sigma_{0} = \left(\frac{1}{2}g \mu_{\mathrm{B}} \right)^2  \frac{e^2 \nu_F}{\left(\lambda k_{\mathrm{F}} \right)^2}$ for brevity.
}
\label{fig:fig2}  
\end{figure}

The anomalous Hall effect \cite{AHE_RMP} as well as LMC \cite{CortijoPRB2016,ZyuzinWSM,Zyuzin2021} are defined by the non-trivial Berry phase of conducting fermions. 
Intrinsic mechanism \cite{AHE_RMP} of the anomalous Hall effect is given by
\begin{align}\label{currentAHE}
{\bf j}_{\mathrm{DWHE}} = e^2
\left[ \int\frac{d {\bf k}}{(2\pi)^2}\sum_{n = \pm} {\bm \Omega}^{(n)}_{\bf k} 
{\cal F}(\epsilon_{{\bf k},n} ) \right]\times {\bf E},
\end{align}
where ${\cal F}(\epsilon)$ is Fermi-Dirac distribution function.
Following the lines of \cite{Zyuzin2020} the Berry curvature 
\begin{align}
&
\Omega^{(\pm)}_{z;\bf k}  =
2\mathrm{Im} \left(\partial_{k_{x}}\Psi^{\dag}_{{\bf k},\pm} \right) \left(\partial_{k_{y}}\Psi_{{\bf k},\pm} \right)
\\
&
=
\mp \frac{\lambda^2}{2\left( \Delta_{\bf k}^2 + \lambda^2 \tilde{k}^2 \right)^{3/2}}
\left( \Delta_{\bf k} - \tilde{k}_{x}\partial_{x}\Delta_{\bf k} - \tilde{k}_{y}\partial_{y}\Delta_{\bf k}\right).
\nonumber
\end{align}
in our model is derived to be
\begin{align}
\Omega^{(\pm)}_{z;\bf k}  = \mp \frac{\lambda^2 \Delta_{\bf k} - \lambda \beta \left( k_{x}h_{x} - k_{y}h_{y} \right)}{2\left(\Delta_{\bf k}^2  + \lambda^2 \tilde{k}^2  \right)^{3/2}}.
\end{align}
It is clear that if $h_{x}=h_{y}=0$ the integral of the Berry curvature over the angles vanishes because of the $d-$wave symmetry. 
Thus, the anomalous Hall effect is absent in this case.
We define $\sigma_{\mathrm{DWHE}}$ as in Eq. (\ref{currentSymmetry}), i.e. as ${\bf j}_{\mathrm{DWHE}} =  \sigma_{\mathrm{DWHE}}\beta B_{x}B_{y}\left[{\bf e}_{z}\times {\bf E}\right]$. 
When $h_{x}\neq 0$ and $h_{y}\neq 0$ the DWHE is non-zero, and we plot it in Fig. (\ref{fig:fig2}).
In addition, we give approximate analytical expressions for various limits of the physical parameters. 
In the limit of $h_{x/y} \ll \lambda k_{\mathrm{F}}$ we have
\begin{align}\label{eq1}
&\sigma_{\mathrm{DWHE}} \approx \sigma_{0} \vert\lambda\vert k_{\mathrm{F}}
\frac{(\beta k_{\mathrm{F}}^2 )^4 + 11 (\beta k_{\mathrm{F}}^2 )^2(\lambda k_{\mathrm{F}})^2 + 16(\lambda k_{\mathrm{F}})^4}{\left[(\beta k_{\mathrm{F}}^2 )^2 + (2\lambda k_{\mathrm{F}})^2 \right]^{5/2}},
\end{align}
where $\nu_{\mathrm{F}} = \frac{m}{\pi}$ is the density of states, $k_{\mathrm{F}} = \sqrt{2m \mu}$ is the Fermi momentum, and where we defined $\sigma_{0} = \left(\frac{1}{2}g \mu_{\mathrm{B}} \right)^2  \frac{e^2 \nu_F}{\left(\lambda k_{\mathrm{F}} \right)^2} $ for brevity.
This dependence is shown in red in Fig. (\ref{fig:fig2}). In the same limit, $h_{x/y} \ll \lambda k_{\mathrm{F}}$, but $\beta k_{\mathrm{F}}^2 < \lambda k_{\mathrm{F}}$ we approximate, 
\begin{align}\label{eq2}
\sigma_{\mathrm{DWHE}} \approx &\sigma_{0}  \bigg[\frac{(\lambda k_F)^2}{(\lambda k_F)^2-h^2} 
+ \frac{1}{16}\frac{(\lambda k_F)^2-8h^2}{(\lambda k_F)^4}\left(\beta k_F^2\right)^2
\nonumber
\\
&
- \frac{1}{64}\frac{6(\lambda k_F)^2 + h^2}{(\lambda k_F)^6}\left(\beta k_F^2\right)^4\bigg].
\end{align}
This dependence is shown in blue in Fig. (\ref{fig:fig2}). When both $h_{x} \gg \beta k^2_{\mathrm{F}}$ and $h_{y} \gg \beta k^2_{\mathrm{F}}$ we approximate
\begin{align}
\sigma_{\mathrm{DWHE}} \approx \sigma_{0}\frac{(\lambda k_{\mathrm{F}})^4}{h^4}, 
\end{align}
where $h^2=h_{x}^2+h_{y}^2$. Thus the magnitude of the corresponding part of the electric current decays with the magnetic field as an inverse square of the field.

%-----------------------------------------------------------------------------------------------------------------
%-----------------------------------------------------------------------------------------------------------------
%-----------------------------------------------------------------------------------------------------------------
\begin{figure}[h] 
\centerline{
\includegraphics[width=0.9 \columnwidth]{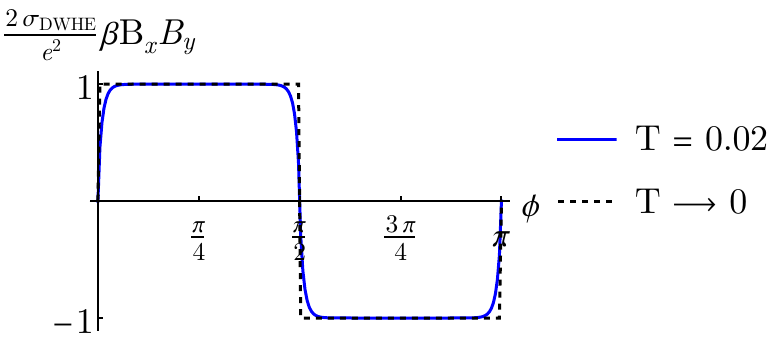} 
}
\protect\caption{
Quantized DWHE in insulating system. 
Insulator is achieved by setting $\frac{k^2}{2m} \rightarrow 0$ and $\mu=0$. 
Plot of the anomalous Hall conductivity $\sigma_{\mathrm{DWHE}}$ as a function of the angle $\phi$ of the in-plane Zeeman magnetic field. Here $h_{x} = h \cos(\phi)$ and $h_{y} = h \sin(\phi)$. Values of the parameters used in the numerical calculation are $h=0.5$, $\mu=0$, $\lambda = 0.5$, and $\beta=1.5$.
}
\label{fig:fig3}  
\end{figure}
%-----------------------------------------------------------------------------------------------------------------
Let us now briefly mention insulating case. We set $\frac{k^2}{2m} \rightarrow 0$ by assuming $m \rightarrow \infty$, and $\mu=0$, then the spectrum (\ref{spectrum}) becomes $\varepsilon_{\bf k}^{(\pm)} = \pm \sqrt{\Delta_{\bf k}^2  + \lambda^2 \tilde{k}^2  }$. By setting $\mu=0$ the system becomes insulating with a gap equal to $2\sqrt{h_{x}^2+h_{y}^2}$ at $k_{x}=k_{y}=0$, and to $2\beta \frac{h_{x}h_{y}}{\lambda^2}$ at $\tilde{k}_{x}=\tilde{k}_{y}=0$. 
In Fig. (\ref{fig:fig3}) we plot $\sigma_{\mathrm{DWHE}}\beta B_{x}B_{y}$, proportionality coefficient in the right hand side of Eq. (\ref{currentAHE}) between the current and the electric field. 
Only the valence band contributes at $T = 0$ to the current. The conductivity is quantized as $\frac{e^2}{2}$ as expected \cite{AHE_RMP,BerryReview}, vanishes when either $h_{x}$ or $h_{y}$ is zero, and changes sign in accord with $d-$wave symmetry.

We note that the predicted DWHE in three-dimensional system will be experimentally measured together with the $\propto ({\bf E}\cdot{\bf B}){\bf B} = \left(E_{x}B_{x}+E_{y}B_{y} \right)\left(B_{x}{\bf e}_{x} + B_{y}{\bf e}_{y} \right)$ term in the electric current. In this term it will appear that $B_{x}B_{y}\left(E_{x}{\bf e}_{y} + E_{y}{\bf e}_{x}\right)$ part is looking like the predicted DWHE, however, the latter is $\propto \beta B_{x}B_{y}\left( E_{x}{\bf e}_{y} - E_{y}{\bf e}_{x} \right)$, and one will have to filter it out from the former. 

Finally, we note that there are other mechanisms contributing to the anomalous Hall effect \cite{AHE_RMP,SinitsynPRB2007,NunnerPRB2007}. They are the skew-scattering and side-jump scattering processes due to the impurities, which are expected to alter the amplitude of the predicted here DWHE but not its symmetry. Their consideration is left for future research.

Let us now discuss linear magnetoconductivity. 
Although, as we have shown above, integral of the Berry curvature over the angles vanishes when $h_{x} = h_{y} = 0$, the Berry curvature can still contribute to the electric current through, for example, modification of the density of states \cite{BerryReview}. 
To study the electric current, we employ the method of kinetic equation,
\begin{align}\label{kineticequation}
\frac{\partial n^{(s)}_{{\bf k}}}{\partial t}
+ {\dot {\bf k}}^{(s)} \frac{\partial n^{(s)}_{{\bf k}}}{\partial {\bf k}}
+ {\dot {\bf r}}^{(s)} \frac{\partial n^{(s)}_{{\bf k}}}{\partial {\bf r}}
= I_{\mathrm{coll}}[ n^{(s)}_{{\bf k}} ],
\end{align}
with equations of motion updated in the presence of the Berry curvature \cite{BerryReview},
$
{\dot {\bf r}}^{(s)} = \frac{\partial \epsilon^{(s)}_{{\bf k}} }{\partial {\bf k}}
+{\dot {\bf k}}^{(s)}\times {\bf \Omega}^{(s)}_{{\bf k}},
$
and 
$
 {\dot {\bf k}}^{(s)} = e{\bf E} + \frac{e}{c}{\dot {\bf r}}^{(s)}\times{\bf B}.
$
The current is given by
$
{\bf j } = e\sum_{s=\pm}\int_{\bf k} \left[ 1 + \frac{e}{c}({\bf B}\cdot {\bf \Omega}^{(s)}_{{\bf k}})\right] \dot{{\bf r}}^{(s)}_{\bf k} n^{(s)}_{\bf k}.
$
We approximate the kinetic equation only by intra-band scattering,
$
I_{\mathrm{coll}}[ n^{(\pm)}_{{\bf k}} ]
= ({\bar n}^{(\pm)} - n^{(\pm)}_{{\bf k}})\tau^{-1},
$
where ${\bar n}^{(s)} = (4\pi)^{-1}\int \sin(\theta)d\theta d\phi \left[ 1 + \frac{e}{c}({\bf B}\cdot {\bf \Omega}^{(s)}_{{\bf k}})\right] n^{(s)}_{\bf k} $ is the distribution function averaged over the angles, and $\tau$ is the fermion's life-time due to the elastic scattering on impurities.
Inter-band scatterings are also allowed but only by virtue of the spin-orbit coupling $\lambda$, since, without it, the bands are spin polarized and there is no scattering between them. 
Then these processes will contribute in higher order in spin-orbit coupling than what we will derive.
Besides, there is no chiral anomaly in the system and, therefore, inter-band scattering processes are not important.

The kinetic equation is approximated as usual, we follow the lines of \cite{ZyuzinWSM,Zyuzin2021} to obtain for the LMC defined as ${\bf j}_{\mathrm{LMC}} = \sigma_{\mathrm{LMC}}\beta B_{z}\left(E_{x}{\bf e}_{y} + E_{y}{\bf e}_{x} \right)$, the following expression 
\begin{align}
\sigma_{\mathrm{LMC}} = - 2 e^2 \nu_F\frac{e\tau}{m c}\frac{\vert\lambda\vert k_{\mathrm{F}}}{\sqrt{(2\lambda k_{\mathrm{F}})^2 + (\beta k_{\mathrm{F}}^2)^2}}.
\end{align}
We note that it is the correction to the density of states \cite{BerryReview} due to the $\frac{e}{c}({\bf B}\cdot {\bf \Omega}^{(s)}_{{\bf k}})$ non-zero product that contributes to this current.
In \cite{Zyuzin2021} it has been shown that linear magnetoconductivity in ferromagnets can be $\delta {\bf j} = \alpha_{1}({\bf E}\cdot{\bf B}){\bf M} + \alpha_{2}({\bf E}\cdot{\bf M}){\bf B} + \alpha_{3}({\bf M}\cdot{\bf B}){\bf E}$, where Onsager relation requires $\alpha_{1}=\alpha_{2}$ (see also comment \cite{comment2}), whose parts were recently experimentally observed in \cite{WSMCorrelated,LeeRosenbaum2020,ExpPRL2021}. Here we found a distinct structure of the LMC. There is a strong evidence that predicted here ${\bf j}_{\mathrm{LMC}}$ has already been observed in \cite{Exp2023} (red arrows in the central figure in Fig. 2 in \cite{Exp2023}). 
There a sign of the voltage drop transverse to the passed current has a $d$-wave symmetry with respect to the direction of the current.  In AHE, given in Eqs. (\ref{AHE}, \ref{IPHE}, \ref{dwave}), the sign does not depend on the direction of the passed current, while Eq. (\ref{lmc}) does and it has the observed $d-$wave symmetry. Indeed, according to Eq. (\ref{lmc}) transverse to the current voltage drop vanishes when the current is passed at $\frac{\pi}{4} + \frac{\pi}{2}n$ angles.

We see that the results decay as a power law in the high density limit. On the other hand, antiferromagnetism does not survive extensive doping of the system with conducting electrons. Therefore, our results are expected to be experimentally observed in the low-doping regime of antiferromagnets. We speculate our predicted DWHE might be relevant to the polar Kerr effect observed in the pseudogap phase of cuprates \cite{cuprateEXP}. 
Indeed, either polar Kerr effect or Faraday rotation is due to the off-diagonal elements of the dielectric tensor, which are defined by the Hall effect in the medium. 
Then the question is which of the anomalous Hall effects, Eq. (\ref{AHE}), Eq. (\ref{IPHE}) or Eq. (\ref{dwave}), contributes.

%-----------------------------------------------------------------------------------------------------------------
%\begin{figure}[h] 
%\centerline{
%\includegraphics[width=0.7 \columnwidth]{iphe2.pdf} 
%}
%\protect\caption{
%Quantized d-wave Hall effect in insulating system. 
%Insulator is achieved by setting $\frac{k^2}{2m} \rightarrow 0$ and $\mu=0$. 
%Plot of the anomalous Hall conductivity $\sigma_{\mathrm{IPHE}}$ as a function of the angle $\phi$ of the in-plane Zeeman magnetic field. Here $h_{x} = h \cos(\phi)$ and $h_{y} = h \sin(\phi)$. Values of the parameters used in the numerical calculation are %$h=0.5$, $\mu=0$, $\lambda = 0.5$, $\beta=1$, and $T=0.01$.
%}
%\label{fig:iphe1}  
%\end{figure}
%-----------------------------------------------------------------------------------------------------------------

\textit{Acknowledgements.}
We thank I.S. Burmistrov, A.M. Finkel'stein, M.M. Glazov, A.S. Mel'nikov, and J. Sinova for helpful discussions.
Both authors are supported by the Foundation for the Advancement of Theoretical Physics and Mathematics BASIS.
VAZ is grateful to Pirinem School of Theoretical Physics.

\begin{widetext}
\newpage
\clearpage
\onecolumngrid
\begin{center}
\rule{0.38\linewidth}{1pt}\\
\vspace{-0.37cm}\rule{0.49\linewidth}{1pt}
\end{center}
\setcounter{section}{0}
\setcounter{equation}{0}
\setcounter{figure}{0}

\section*{Supplemental Material to  "d-Wave Hall effect and linear magnetoconductivity in metallic collinear antiferromagnets"}

Here we present symmetry argument used in the Main Text to understand the structure of electric current in the studied system.
We assume that the unperturbed system is just a square lattice, which obeys the $D_{4h}$ symmetry. 
Terms $\beta\sigma_{z}k_{x}k_{y}$, $h_{x/y}\sigma_{x/y}$, orbital part of the $B_{z}$ magnetic field, $\lambda (\sigma_{x}k_{y} - \sigma_{y}k_{x})$, and $E_{x/y}$ in the system are considered in our symmetry argument as perturbations.

Definitions and elements multiplication table are given in many resources. 
We have used tables given in the Professor D.W. Snoke website [1SM], pages 169-180. 
References [2SM, 3SM] and many more can be of use.

In the given group $D_{4h}$ these perturbations transform with the following irreducible representations,
$B_{x}$ as $\Gamma_{5}^{(+)}$ with $\vert 1 \rangle$, $B_{y}$ as $\Gamma_{5}^{(+)}$ with $\vert 2 \rangle$. 
Orbital part of the $B_{z}$ as $\Gamma_{2}^{(+)}$ with $\vert 1 \rangle$,
combination $\sigma_{z}k_{x}k_{y}$ as well as $\beta$ changes as $\Gamma_3^{(+)}$ with $\vert 1 \rangle$, 
Rashba spin-orbit coupling as $\lambda$ as well as $k_{x}\sigma_{y} - k_{y}\sigma_{x}$ transform as $\Gamma_{2}^{(-)}$ with $\vert 1 \rangle$,
and the electric field $E_{x}$ as $\Gamma_{5}^{(-)}$ with $\vert 1 \rangle$ and $E_{y}$ as $\Gamma_{5}^{(-)}$ with $\vert 2 \rangle$.

We note that it is the fields, $B_{x/y}$, $B_{z}$, $\beta$, $\lambda$, and $E_{x/y}$ that transform that way. 
Whatever these fields couple to, which may be called as the charges, for example $\beta$ couples to $\sigma_{z}k_{x}k_{y}$, $B_{x/y}$ couples to $\sigma_{x/y}$ transform according to the tables given in [1SM].
For example, in the $\sigma_{z}k_{x}k_{y}$ term $\sigma_{z}$ transforms as $\Gamma_{2}^{(+)}$ with $\vert 1 \rangle$, and $k_{x}k_{y}$ as $\Gamma_{4}^{(+)}$ with $\vert 1 \rangle$. 
Then the whole $\sigma_{z}k_{x}k_{y}$ term, according to the multiplication table in [1SM], transforms as $\Gamma_{4}^{(+)}\times\Gamma_{2}^{(+)} = \Gamma_{3}^{(+)}$. 
Therefore, $\beta$ transforms in the same way because $\Gamma_{3}^{+}\times\Gamma_{3}^{+} = \Gamma_{1}^{+}$ is the identity. 

Let us show what element the Rashba spin-orbit coupling transforms with. Momentum $k_{x/y}$ transforms with $\Gamma_{5}^{(-)}$ with $\vert 1/(2) \rangle$, while $\sigma_{x/y}$ transforms with $\Gamma_{5}^{(+)}$ with $\vert 1/2 \rangle$ (where $1/2$ is not a fraction but cooresponding labeling of $x/y$).
We will shorten the notations from $\Gamma_{5}^{(-)}$ with $\vert 1 /2 \rangle$ and $\Gamma_{5}^{(+)}$ with $\vert 1 / 2 \rangle$ to $\Gamma_{5}^{(-)}\vert 1 / 2 \rangle$ and $\Gamma_{5}^{(+)}\vert 1 / 2 \rangle$ correspondingly for the sake of brevity.
Then $k_{x}\sigma_{y}$, according to the multiplication table shown in [1SM], transforms as 
\begin{align}
\Gamma_{5}^{(-)}\vert 1 \rangle \times \Gamma_{5}^{(+)}\vert 2 \rangle = \frac{1}{\sqrt{2}}\Gamma_{2}^{(-)}\vert 1 \rangle + \frac{1}{\sqrt{2}}\Gamma_{4}^{(-)}\vert 1 \rangle,
\end{align} 
while $k_{y}\sigma_{x}$ as 
\begin{align}
\Gamma_{5}^{(-)}\vert 2 \rangle \times \Gamma_{5}^{(+)}\vert 1 \rangle =- \frac{1}{\sqrt{2}}\Gamma_{2}^{(-)}\vert 1 \rangle + \frac{1}{\sqrt{2}}\Gamma_{4}^{(-)}\vert 1 \rangle.
\end{align}
Therefore, $k_{x}\sigma_{y} - k_{y}\sigma_{x}$ combination transforms as $\Gamma_{2}^{(-)}\vert 1 \rangle $, namely 
\begin{align}
k_{x}\sigma_{y} - k_{y}\sigma_{x} :~~~
\frac{1}{\sqrt{2}}\Gamma_{2}^{(-)}\vert 1 \rangle + \frac{1}{\sqrt{2}}\Gamma_{4}^{(-)}\vert 1 \rangle + \frac{1}{\sqrt{2}}\Gamma_{2}^{(-)}\vert 1 \rangle - \frac{1}{\sqrt{2}}\Gamma_{4}^{(-)}\vert 1 \rangle
\rightarrow \Gamma_{2}^{(-)}\vert 1 \rangle, 
\end{align}
where by second rightarrow we disregarded the coefficient of $\sqrt{2}$, as we are interested only in the element.
Finally, again using the multiplication table in [1SM], we see that $\lambda$ field transforms as $\Gamma_{2}^{-}\vert 1 \rangle$ as well.

We are interested in linear response of the electric current to the electric field $E_{x,y}$. 
The electric current $j_{x/y}$ transforms as $\Gamma_{5}^{(-)}$ with $\vert 1 / 2 \rangle$, where $\vert 1 / 2 \rangle$ correspond to $x/y$ projection of the electric current.  
Therefore, we must find all combinations of the perturbations to linear order in $\beta$, $E_{x/y}$, $B_{z}$, quadratic in $B_{x/y}$ and to lowest possible order in $\lambda$ whose product results in $\Gamma_{5}^{(-)}$. 
We stress that it is either the fields, $\beta\sigma_{z}k_{x}k_{y}$, $B_{x/y}\sigma_{x/y}$, orbital part of the 
 $B_{z}$ magnetic field, $\lambda (\sigma_{x}k_{y} - \sigma_{y}k_{x})$, and $E_{x/y}$, we have to multiply or their correponding charges. 
Indeed, on the macroscopic level, when traces over the momenta and spin are performed, we will be left with the product of the fields. 
Therefore, it is intuitive that we could have multipled only the fields without thinking how the charges multiply.

Calculations show that $B_{x}^2 - B_{y}^2$ and $B_{x} B_{y} B_{z}$ have representation $\Gamma_3^{(+)}$, and $\beta B_z$ and $B_x B_y$ have representation $\Gamma_4^{(+)}$ and finally $\beta B_x B_y$ has a representation $\Gamma_2^{(+)}$. 
It is convenient to list all invariants  to lowest order in fields, which are $\beta^2$, $\lambda^2$, $B_z^2$, $B_x^2 + B_y^2$ and $\beta(B_x^2 - B_y^2)$. 
They can couple to electric field and thus result in electric current if in addition they obey the Onsager relation for the conductivity tensor, which reads as $\sigma_{ij} ({\bf B}, \beta) = \sigma_{ji}(-{\bf B}, -\beta)$.

In addition, it is important to note that odd powers of $\lambda$ don't enter the expression for the electric current. 
This is because $\lambda$ transforms as $\Gamma_{2}^{-}$ and according to the rules of the parity multiplication, which are $-\times - = +$, $- \times + = -$, $+ \times - = - $ and $+ \times + = +$, and since the only other negaitve parity field available in the system is the electric field, we will always end up with the $+$ parity for the overall product, while the electric current has $-$ parity. Hence, there is no way to obtain electric current in linear or any odd power of $\lambda$.

We take $\beta B_x B_y$ which transforms as $\Gamma_2^{(+)}$ with element $\vert 1\rangle$ and fuse it with the electric field $E_{x}$ and $E_{y}$ which transform as $\Gamma_5^{(-)}$ with basis vectors $\vert 1 \rangle$ and $\vert 2 \rangle$ correspondingly, and get
\begin{align}
&
\Gamma_2^{(+)}\vert 1\rangle \times \Gamma_{5}^{(-)}\vert 1 \rangle   = -  \Gamma_{5}^{(-)}\vert 2 \rangle, \\
&
\Gamma_2^{(+)}\vert 1\rangle \times \Gamma_{5}^{(-)}\vert 2 \rangle   =   \Gamma_{5}^{(-)}\vert 1 \rangle.
\end{align}
Therefore, $\delta j_{x} \propto \beta B_x B_y E_{y}$ and $\delta j_{y} \propto - \beta B_x B_y E_{x}$, which compactly reads as $\delta {\bf j} \propto \beta B_x B_y \left[{\bf e} \times {\bf E}\right]_{z} $. 

Now we take $\beta B_z$ and/or $B_{x}B_{y}$ which transforms as $\Gamma_4^{(+)}$ with basis vector $\vert 1 \rangle$ and fuse it with $E_{x}$ and $E_{y}$, and obtain
\begin{align}
&
\Gamma_4^{(+)}\vert 1\rangle \times \Gamma_{5}^{(-)}\vert 1 \rangle   =   \Gamma_{5}^{(-)}\vert 2 \rangle, \\
&
\Gamma_4^{(+)}\vert 1\rangle \times \Gamma_{5}^{(-)}\vert 2 \rangle   =   \Gamma_{5}^{(-)}\vert 1 \rangle.
\end{align}
Therefore, $\delta {\bf j} \propto \beta B_z \left( E_{x}{\bf e}_{y} + E_{y}{\bf e}_{x} \right)$ and $\delta {\bf j} \propto B_{x}B_{y} \left( E_{x}{\bf e}_{y} + E_{y}{\bf e}_{x} \right)$.

Finally, let us take for example $B_{x}^2 - B_{y}^2$ combination which transforms as $\Gamma_3^{(+)}$ with vector $\vert 1 \rangle$ and fuse it again with $E_{x}$ and $E_{y}$,
 \begin{align}
&
\Gamma_3^{(+)}\vert 1\rangle \times \Gamma_{5}^{(-)}\vert 1 \rangle   =   \Gamma_{5}^{(-)}\vert 1 \rangle, \\
&
\Gamma_3^{(+)}\vert 1\rangle \times \Gamma_{5}^{(-)}\vert 2 \rangle   =   - \Gamma_{5}^{(-)}\vert 2 \rangle.
\end{align}
Therefore, $\delta{\bf j} \propto \left( B_{x}^2 - B_{y}^2 \right)\left(E_{x}{\bf e}_{x} - E_{y}{\bf e}_{y} \right)$. 

Finally, since $\beta(B_x^2 - B_y^2)$ is an invariant, then one may think that there is a term $\delta{\bf j} \propto \beta(B_x^2 - B_y^2){\bf E}$ in the current. 
However, this term does not obey Onsager reciprocity relation, because $\beta \rightarrow - \beta$ under the time-reversal operation. Possible $\delta{\bf j} \propto \beta(B_x^2 - B_y^2)[{\bf e}_{z}\times {\bf E}]$ term in the current which would obey the Onsager relation can't be created under the fusion of the fields, hence, it does not exist.

Collecting all the terms we get for the current,
\begin{align}
{\bf j} 
&
= \sigma_{\mathrm{D}} {\bf E} + \sigma_{\mathrm{H}} \left[{\bf E}\times {\bf B} \right] + \sigma_{2A}H_{x}H_{y} \left( E_{x}{\bf e}_{y} + E_{y}{\bf e}_{x} \right) + \sigma_{2B} \left( H_{x}^2 - H_{y}^2 \right)\left(E_{x}{\bf e}_{x} - E_{y}{\bf e}_{y} \right)
\\
&
 + \sigma_{\mathrm{LMC}}\beta B_{z}\left(E_{x}{\bf e}_{y} + E_{y}{\bf e}_{x} \right) + \sigma_{\mathrm{DWHE}}\beta B_{x}B_{y}\left[{\bf e}_{z}\times {\bf E}\right].
\end{align}
In isotropic system $\sigma_{2A} = \sigma_{2B} \equiv \sigma_{2}$, and as a result of it the current reads
\begin{align}
{\bf j} 
= \sigma_{\mathrm{D}} {\bf E} + \sigma_{\mathrm{H}} \left[{\bf E}\times {\bf B} \right] + \sigma_{2}\left[  ({\bf E}\cdot{\bf B} ) {\bf B} - {\bf B}^2 {\bf E} \right]
 + \sigma_{\mathrm{LMC}}\beta B_{z}\left(E_{x}{\bf e}_{y} + E_{y}{\bf e}_{x} \right) + \sigma_{\mathrm{DWHE}}\beta B_{x}B_{y}\left[{\bf e}_{z}\times {\bf E}\right].
\end{align}
The first three terms here are consistent with Ref. [4SM]. This expression reproduces corresponding equation in the Main Text.

\subsection{References in Supplemental Material }
\begin{enumerate}

\item https://www.snokelab.com/symmetry-tables

\item G.F. Koster, J.O. Dimmock, R.G. Wheeler, and H. Statz, \textit{Properties of The Thirty-Two Point Groups}, MIT Press, Cambridge MA, (1963).

\item G.L. Bir and G.E. Pikus, \textit{Symmetry and Strain-induced Effects in Semiconductors}, Wiley 1974.

\item F. Seitz, Phys. Rev. {\bf 79}, 372 (1950), \textit{Note on the theory of resistance of a cubic semiconductor in a magnetic field}.

\end{enumerate}

\end{widetext}


\begin{references}

%--------------------------------------Introduction

\bibitem{KarplusLuttinger} R. Karplus and J.M. Luttinger, Phys. Rev. {\bf 95}, 1154 (1954)
\textit{Hall Effect in Ferromagnetics}

\bibitem{Hall} E.H. Hall, Philos. Mag.{\bf 10}, 301 (1880).
\textit{On the new action of magnetism on a permanent electric current}

\bibitem{Ziman} J. M. Ziman,\textit{ Electrons and Phonons: The Theory of Transport Phenomena in Solids}, (Oxford
University Press, Oxford, U.K., 1960).

\bibitem{Vas'ko} F.T. Vas'ko, JETP Lett. {\bf 30}, 541 (1979); 
\textit{Spin splitting in the spectrum of two-dimensional electrons due to the surface potential}

\bibitem{BychkovRashba} Yu. A. Bychkov and E. I. Rashba, JETP Lett. {\bf 39}, 78 (1984).
\textit{Properties of a 2D electron gas with lifted spectral degeneracy}

\bibitem{Dresselhaus} G. Dresselhaus, Phys. Rev. {\bf 100}, 580 (1955).
\textit{Spin-orbit coupling effects in zinc blende structures}






\bibitem{SovietTI1} B. A. Volkov and O. A. Pankratov, Pis’ma Zh. Eksp. Teor. Fiz. {\bf 42}, 145 (1985) [JETP Lett. {\bf 42}, 178 (1985)].
\textit{Two-dimensional massless electrons in an inverted contact}

\bibitem{SovietTI2}
 O.A. Pankratov, S.V. Pakhomov, and B.A. Volkov, Solid State Communications, {\bf 61}, 93 (1987).
\textit{Supersymmetry in heterojunctions: Band-inverting contact on the basis of Pb1-xSnxTe and Hg1-xCdxTe}



\bibitem{Dyakonov} \textit{Spin Physics in Semiconductors}, edited by M. I. Dyakonov
(Springer-Verlag, Berlin, Heidelberg, 2008).



\bibitem{AHE_RMP} N. Nagaosa, J. Sinova, S. Onoda, A. H. MacDonald, and N. P. Ong, Rev. Mod. Phys. {\bf 82}, 1539 (2010).
\textit{Anomalous Hall effect}

\bibitem{CulcerMacDonaldNiuPRB2003} D. Culcer, A. H. MacDonald, and Q. Niu, Phys. Rev. B {\bf 68}, 045327 (2003).
\textit{Anomalous Hall effect in paramagnetic two-dimensional systems}




\bibitem{SinitsynPRB2007} N. A. Sinitsyn, A. H. MacDonald, T. Jungwirth, V. K. Dugaev, and J. Sinova, Phys. Rev. B {\bf 75}, 045315 (2007).
\textit{Anomalous Hall effect in a two-dimensional Dirac band: The link between the Kubo-Streda formula and the semiclassical Boltzmann equation approach}


\bibitem{NunnerPRB2007} T. S. Nunner, N. A. Sinitsyn, M. F. Borunda, V. K. Dugaev, A. A. Kovalev, Ar. Abanov, C. Timm, T. Jungwirth, J. Inoue, A. H. MacDonald, and J. Sinova, Phys. Rev. B {\bf 76}, 235312 (2007).
\textit{Anomalous Hall effect in a two-dimensional electron gas}


\bibitem{Berry} M. Berry, Proc. R. Soc. London, Ser. A {\bf 392}, 45 (1984).
\textit{Quantal phase factors accompanying adiabatic changes}



\bibitem{BerryReview} D. Xiao, M.C. Chang, and Q. Niu, Rev. Mod. Phys. {\bf 82}, 1959 (2010).
\textit{Berry phase effects on electronic properties}

%-------------------------------------in-plane Hall effect



\bibitem{Malshukov1998} A.G. Mal'shukov, K.A. Chao, and M. Willander, Phys. Rev. B {\bf 57} 2069(R) (1998).
\textit{Hall effect in a magnetic field parallel to interfaces of a III-V semiconductor quantum well}


\bibitem{LiuPRL2013} X. Liu, H.C. Hsu, and C.X. Liu, Phys. Rev. Lett. {\bf 111}, 086802 (2013).
\textit{In-Plane Magnetization-Induced Quantum Anomalous Hall Effect}


\bibitem{ZhangPRB2019} J. Zhang, Z. Liu, and J. Wang, Phys. Rev. B {\bf 100}, 165117 (2019).
\textit{In-plane magnetic-field-induced quantum anomalous Hall plateau transition}


\bibitem{Zyuzin2020} V.A. Zyuzin, Phys. Rev. B {\bf 102}, 241105(R) (2020).
\textit{In-plane Hall effect in two-dimensional helical electron systems}

\bibitem{Culcer2021} J.H. Cullen, P. Bhalla, E. Marcellina, A.R. Hamilton, and D. Culcer, Phys. Rev. Lett. {\bf 126}, 256601 (2021).
\textit{Generating a Topological Anomalous Hall Effect in a Nonmagnetic Conductor: An In-Plane Magnetic Field as a Direct Probe of the Berry Curvature}

\bibitem{Kurumaji2023} T. Kurumaji, arXiv:2304.00785 (2023).
\textit{Symmetry-based requirement for the measurement of electrical and thermal Hall conductivity under an in-plane magnetic field}



\bibitem{AHE_ZrTe5} T. Liang, J. Lin, Q. Gibson, S. Kushwaha, M. Liu, W. Wang,
H. Xiong, J.A. Sobota, M. Hashimoto, P.S. Kirchmann, Z.-X. Shen,
R.J. Cava, and N.P. Ong, Nature Physics {\bf 14}, 451 (2018).
\textit{Anomalous Hall effect in ZrTe$_5$} 


\bibitem{Zhang_IPHE2020} C.L. Zhang, T. Liang, N. Ogawa, Y. Kaneko, M. Kriener, T. Nakajima,
Y. Taguchi, and Y. Tokura, Phys. Rev. Mat. {\bf 4}, 091201(R) (2020).
\textit{Highly tunable topological system based on PbTe-SnTe binary alloy}


\bibitem{Liu2024_C3v} L. Liu, A. Pezo, D.G. Ovalle, C. Zhou, Q. Shen, H. Chen,
T. Zhao, W. Lin, L. Jia, Q. Zhang, H. Zhou, Y. Yang, A. Manchon,
and J. Chen, Nano Letters {\bf 24}, 733 (2024).
\textit{Crystal symmetry-dependent in-plane Hall effect}


%---------------------------------------------------- AFM spin-splitting  --------------------


\bibitem{AHE_AFM_Review} L. Šmejkal, A.H. Macdonald, J. Sinova, S. Nakatsuji, and T. Jungwirth, Nature Review Materials {\bf 7}, 482 (2022).
\textit{Anomalous Hall antiferromagnets}

\bibitem{AHE_AFM} L. Šmejkal, R. González-Hernández, T. Jungwirth, and J. Sinova., Science Advances {\bf 6}, eaaz8809 (2020).
\textit{Crystal time-reversal symmetry  breaking and spontaneous Hall effect in collinear antiferromagnets}



\bibitem{PekarRashba} S.I. Pekar and E.I. Rashba, Sov. Phys. JETP {\bf 20}, 1295 (1965).
\textit{Combined resonance in crystals in inhomogeneous magnetic fields}



\bibitem{Ogg1966} N.R. Ogg, Proc. Phys. Soc. {\bf 89}, 431 (1966).
\textit{Conduction-band g factor anisotropy in indium antimonide}

\bibitem{IvchenkoKiselevPTS1992} E.L. Ivchenko and A.A. Kiselev, PTS {\bf 26}, 1471 (1992).





\bibitem{VarmaZhu2006} C.M. Varma and L. Zhu, Phys. Rev. Lett. {\bf 96}, 036405 (2006).
\textit{Helicity Order: Hidden Order Parameter in URu$_2$Si$_2$}

\bibitem{WuSunFradkinZhang2007} C. Wu, K. Sun, E. Fradkin, and S.-C. Zhang, Phys. Rev. B {\bf 75}, 115103 (2007).
\textit{Fermi liquid instabilities in the spin channel}

\bibitem{Ramazashvili} R. Ramazashvili Phys. Rev. B {\bf 79}, 184432 (2009).
\textit{Kramers degeneracy in a magnetic field and Zeeman spin-orbit coupling in antiferromagnetic conductors}

\bibitem{HayamiYanagiKusunose2019} S. Hayami, Y. Yanagi, and H. Kusunose, J. Phys. Soc. Jpn. {\bf 88}, 123702 (2019).
\textit{Momentum-Dependent Spin Splitting by Collinear Antiferromagnetic Ordering}

\bibitem{Rashba2020} L.-D. Yuan, Z. Wang, J.-W. Luo, E.I. Rashba, A. Zunger, Phys. Rev. B {\bf 102}, 144422 (2020).
\textit{Giant momentum-dependent spin splitting in centrosymmetric low-Z antiferromagnets}

\bibitem{HayamiYanagiKusunose2020} S. Hayami, Y. Yanagi, and H. Kusunose, Phys. Rev. B {\bf 102}, 144441 (2020).
\textit{Bottom-up design of spin-split and reshaped electronic band structures in antiferromagnets without spin-orbit coupling: Procedure on the basis of augmented multipoles}

\bibitem{EgorovEvarestov} S.A. Egorov and R.A. Evarestov, J. Phys. Chem. Lett. {\bf 12}, 2363 (2021).
\textit{Colossal Spin Splitting in the Monolayer of the Collinear Antiferromagnet MnF2}

\bibitem{SmejkalSinovaJungwirth2022a} L. Šmejkal, J. Sinova, and T. Jungwirth, Phys. Rev. X {\bf 12}, 031042 (2022); 
\textit{Beyond Conventional Ferromagnetism and Antiferromagnetism: A Phase with Nonrelativistic Spin and Crystal Rotation Symmetry}


\bibitem{SmejkalSinovaJungwirth2022b} L. Šmejkal, J. Sinova, and T. Jungwirth, Phys. Rev. X {\bf 12}, 040501 (2022).
\textit{Emerging Research Landscape of Altermagnetism}


\bibitem{Bose2022} A. Bose, N.J. Schreiber, R. Jain, D.-F. Shao, H.P. Nair, J. Sun, X.S. Zhang, D.A. Muller, E.Y. Tsymbal, D.G. Schlom, and D.C. Ralph, Nature Electronics {\bf 5}, 267 (2022).
\textit{Tilted spin current generated by the collinear antiferromagnet ruthenium dioxide}

\bibitem{Gonzalez-Hernandez2021} R. González-Hernández, L. Šmejkal, K. Výborný, Y. Yahagi,
J. Sinova, T. Jungwirth, and J. Železný, Phys. Rev. Lett. {\bf 126}, 127701 (2021).
\textit{Efficient Electrical Spin Splitter Based on Nonrelativistic Collinear Antiferromagnetism}


\bibitem{Exp2023} H. Koizumi, Y. Yamasaki, and H. Yanagihara, Nature Communications {\bf 14}, 8074 (2023).
\textit{Quadrupole anomalous Hall effect in magnetically induced electron nematic state} 


\bibitem{Guo_npj2023} P.J. Guo, Z.X. Liu, and Z.Yi. Lu, npj Computational Materials {\bf 9}, 70 (2023). 
\textit{Quantum anomalous hall effect in collinear antiferromagnetism}



%\bibitem{comment1} It has been recently suggested to call such materials as the altermagnets. 
%Here we restrained ourselves from using this name since the system studied in the present work is a collinear antiferromagnet. 


%----------------------------symmetry

\bibitem{Koster} G.F. Koster, J.O. Dimmock, R.G. Wheeler, and H. Statz, \textit{Properties of The Thirty-Two Point Groups}, MIT Press, Cambridge MA, (1963).

\bibitem{BirPikus} G.L. Bir and G.E. Pikus, \textit{Symmetry and Strain-induced Effects in Semiconductors}, Wiley 1974.

\bibitem{Snoke} https://www.snokelab.com/symmetry-tables

%------------------------------------ (EB)B


\bibitem{SeitzPR1950} F. Seitz, Phys. Rev. {\bf 79}, 372 (1950).
\textit{Note on the theory of resistance of a cubic semiconductor in a magnetic field}


\bibitem{GoldbergDavisPR1954} C. Goldberg and R.E. Davis, Phys. Rev. {\bf 94}, 1121 (1954).
\textit{New galvanomagnetic effect}


\bibitem{ZyuzinWSM} V.A. Zyuzin, Phys. Rev. B {\bf 95}, 245128 (2017).
\textit{Magnetotransport of Weyl semimetals due to the chiral anomaly}

\bibitem{KyJETP1966} V.D. Ky, Sov. Phys. JETP {\bf 23}, 809 (1966).
\textit{Plane Hall effect in ferromagnetic metals}


%-----------------------------------linear magnetoconductivity

\bibitem{CortijoPRB2016} A. Cortijo, Phys. Rev. B {\bf 94}, 241105(R) (2016).
\textit{Linear magnetochiral effect in Weyl semimetals}


\bibitem{Zyuzin2021} V.A. Zyuzin, Phys. Rev. B {\bf 104}, L140407 (2021).
\textit{Linear magnetoconductivity in magnetic metals}

\bibitem{comment2} In \cite{Zyuzin2021} Onsager reciprocity relation is not satisfied by the terms in the conductivity which are due to the chiral anomaly. Namely, coefficients in $\delta{\bf j} \propto \alpha_{1}({\bf E}\cdot{\bf B}){\bf M}+\alpha_{2}({\bf E}\cdot{\bf M}){\bf B}$ Onsager relation dictates $\alpha_{1}=\alpha_{2}$, while in \cite{Zyuzin2021} $\alpha_{1}\neq\alpha_{2}$. It is possible, side-jump scattering processes have to be included in order to restore the Onsager reciprocity relation, or it might as well be that the relation does not survive the chiral anomaly, when different chemical potentials in the opposite valleys are created thus driving the system out of equilibirum.

\bibitem{WSMCorrelated} Kuroda, K., Tomita, T., Suzuki, MT. et al., Nature Material {\bf 16}, 1090 (2017).
\textit{Evidence for magnetic Weyl fermions in a correlated metal}

\bibitem{LeeRosenbaum2020} Y. Wang, P.A. Lee, D.M. Silevitch, F. Gomez, S.E. Cooper, Y. Ren, J.-Q. Yan, D. Mandrus, T.F. Rosenbaum, and Y. Feng, Nature Communications {\bf 11}, 216 (2020).
\textit{Antisymmetric linear magnetoresistance and the planar Hall effect}


\bibitem{ExpPRL2021} B. Jiang, L. Wang, R. Bi, J. Fan, J. Zhao, D. Yu, Z. Li, and X. Wu,
Physical Review Letters {\bf 126}, 236601 (2021).
\textit{Chirality-Dependent Hall Effect and Antisymmetric Magnetoresistance in a Magnetic Weyl Semimetal}




%---------------------------------------------------SM
\bibitem{SM} See Supplemental Material at URL for more details.

%------------------------------------ cuprate

\bibitem{cuprateEXP} J. Xia, E. Schemm, G. Deutscher, S.A. Kivelson, D.A. Bonn, W.N. Hardy, R. Liang, W. Siemons,
G. Koster, M.M. Fejer, and A. Kapitulnik, Phys. Rev. Lett. {\bf 100}, 127002 (2008).
\textit{Polar Kerr-Effect Measurements of the High-Temperature YBa$_2$Cu$_3$O6$_x$ Superconductor: Evidence for Broken Symmetry near the Pseudogap Temperature}




\end{references}
\end{document}